\documentclass[times,twocolumn,final,authoryear]{elsarticle}

\usepackage{jasr}
\usepackage{framed,multirow}
\usepackage[left]{lineno}

\usepackage{amssymb}
\usepackage{amsmath}
\usepackage{latexsym}
\usepackage{graphicx}

\usepackage{epstopdf}

\usepackage{color}

\usepackage{amssymb}

\def\lsim{\;\raise0.3ex\hbox{$<$\kern-0.75em\raise-1.1ex\hbox{$\sim$}}\;}
\def\gsim{\;\raise0.3ex\hbox{$>$\kern-0.75em\raise-1.1ex\hbox{$\sim$}}\;}

\newcommand{\ergs}{\,erg\,s$^{-1}$\,}

\usepackage{url}
\usepackage{xcolor}
\definecolor{newcolor}{rgb}{.8,.349,.1}

\usepackage[citebordercolor=white]{hyperref}

\journal{Advances in Space Research}

\begin{document}

\verso{V.I.Romansky et al \textit{etal}}

\begin{frontmatter}

\title{Cosmic ray acceleration and non-thermal emission from fast luminous optical transient sources}%

\author[1]{V.I. {Romansky}\corref{cor1}}
\cortext[cor1]{Corresponding author}
  \ead{romanskyvadim@gmail.com}
\author[1]{A.M. {Bykov}}
\ead{byk@astro.ioffe.ru}
\author[1]{S.M. {Osipov}}
\ead{osm2004@mail.ru}

\address[1]{Ioffe Institute, Saint Petersburg, Polytechnicheskaya str., 26, 194021, Russia}

\received{}
\finalform{}
\accepted{}
\availableonline{}
\communicated{}

\begin{abstract} Fast blue optical transients (FBOTs) represent a new class of highly energetic sources observed from radio to X-rays. High luminosity,  
light curves and spectra of the sources can be understood if they are associated with 
supernova-like or tidal disruption events. Radio observations of the 
transient sources revealed a mildly relativistic expansion of some of the remnants.
The high power and mildly relativistic shock velocities are providing favorable conditions for very high energy particle acceleration. In this paper we present a model of particle acceleration in mildly relativistic magnetohydrodynamic (MHD) outflow of the transient source. To construct the non-thermal radiation and cosmic ray spectra in a broad range of energies we combined the microscopic particle-in-cell (PIC) simulations of electron and proton injection at mildly relativistic shock with Monte Carlo technique for high energy particle transport and acceleration. The kinetic PIC simulations provided the energy partition parameter $\epsilon_{e}$ used to fit the observed non-thermal radio emission using the magnetic field amplification mechanisms modelled with Monte Carlo simulations. The model allowed to describe the radio-spectrum of CSS161010 and it's X-ray luminosity. The high X-ray luminosity of  AT2018 and AT2020mrf detected during the first weeks can be connected to the jet interaction with the stellar companion in a binary system.
The model predicts that FBOTs can accelerate cosmic rays to energies above 10 PeV with a possible upper limit of maximum energy of 100 PeV. With the expected event rate of FBOTs they can contribute to the very high energy cosmic rays population in galaxies.   
\end{abstract}


\end{frontmatter}

\section{Introduction}
Wide diversity of the observed powerful rapidly evolving sources includes the growing population of fast blue optical transients which demonstrate properties different from the known types of supernovae \citep{Drout2014,Tanaka16,Margutti2019,Ho2019cow,Ho2023flares,Coppejans2020,YaoAt2020mrf}. Many of the sources are associated with star-forming galaxies and some preliminary estimates indicate that the event rate can be about a percent of the core-collapsed supernova rate \citep[see e.g.][]{Tanaka16,Margutti2019}.       
The recently discovered FBOTs are highly energetic objects with a very fast rise (a day time scale). They are detected in optical, radio and X-ray band with the peak luminosity reaching in some cases $\sim$ 10$^{44}$ \ergs. Extraordinary high level of the X-ray radiation was also found for the most of FBOTs
(see \cite{Margutti2019, Ho2019cow,Ho2020koala,Coppejans2020, Ho2021at2020, YaoAt2020mrf, MatthewsAT2022tsd, CHrimesAT2023fhn}). Radio and optical observations of FBOTs revealed the presence of fast plasma outflows with mildly relativistic velocities $\gsim 0.1 c$ and total energies of 10$^{50}$ - 10$^{51}$ ergs.  

Different scenarios were proposed to explain the origins of the FBOT radiation including the central engine activity and  the interaction of fast outflows and shocks with external medium \citep[see e.g.][]{Leung_Blinnikov20,Urvachev21,GottliebFBOTjet,Metzger22}. Models of the most luminous FBOTs can be based on the tidal disruption of Wolf-Rayet stars in binaries with a black hole or neutron star companion \citep[][]{Metzger22}. Relativistic jets launched by the central engine of collapsed  hydrogen-rich stars, which are interacting with the circumstellar medium, were proposed to explain the variety of the observed properties of FBOTs. The presence of the energetic mildly relativistic ejecta can be important for acceleration of cosmic rays (CRs) to high energies and production of non-thermal emission. The observed X-ray non-thermal emission with  luminosity $\sim$ 10$^{39}$ \ergs in three months after the event may be used to constrain  models of CR acceleration.  

In this work we consider a model of relativistic particle acceleration in FBOT-type objects. The object CSS161010, described in \cite{Coppejans2020}, has a very high ejecta velocity estimated as $0.5~c$ at first few months. The mildly relativistic shocks, which are likely associated with FBOTs, are favorable for very high energy CR acceleration and can be considered as possible sources of CRs with energies well above the observed in most of the young galactic supernova remnants \citep{2018SSRv..214...41B}. 

 The high shock velocity makes the particle-in-cell technique suitable for modeling of the non-thermal emission. To construct a model of non-thermal emission from FBOT we performed PIC modeling of particle acceleration in mildly relativistic shocks, which can describe the ion injection processes. This allowed us to obtain the distribution function of emitting electrons. 

The paper is organized as follows: in section \ref{particlesChapter} we present the results of PIC and Monte Carlo modeling of electron and CR proton acceleration in mildly-relativistic shocks.  In section \ref{synchrotronChapter} we calculate synchrotron radio flux from CSS161010 using obtained non-thermal particles distributions and fit it to observation data. The synchrotron X-ray flux and the non-thermal broad band spectra are presented in section \ref{compton}. In section \ref{jetChapter} we discuss the alternative model of powerful X-ray radiation from the source via inverse Compton scattering of accelerated electrons in the presence of high-luminous neighbour star.

\section{Particle acceleration by mildly relativistic shock}\label{particlesChapter}
\begin{figure}[h]
	\centering
		\includegraphics[width=0.4\textwidth]{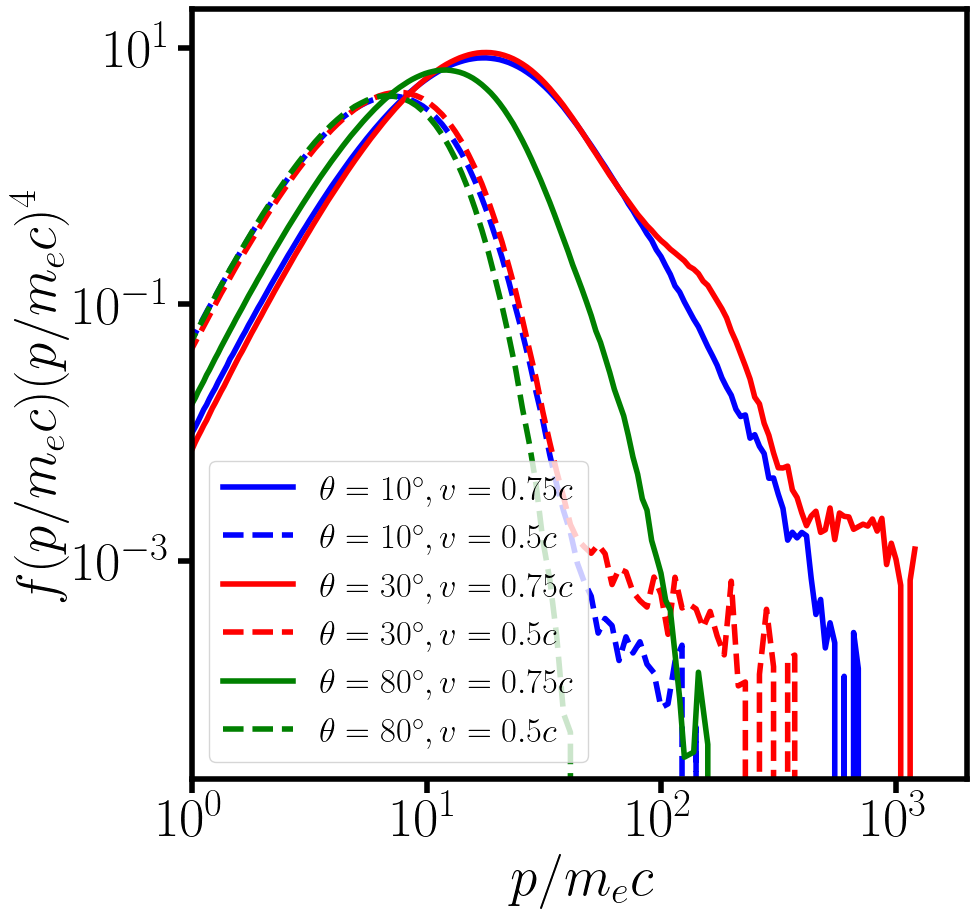} 
		\caption{Electron distribution functions in the downstream of the shocks with different inclination angles $\theta$ and velocities $v$}
		\label{distributionsE}
\end{figure}

\begin{figure}[h]
		\centering
		\includegraphics[width=0.4\textwidth]{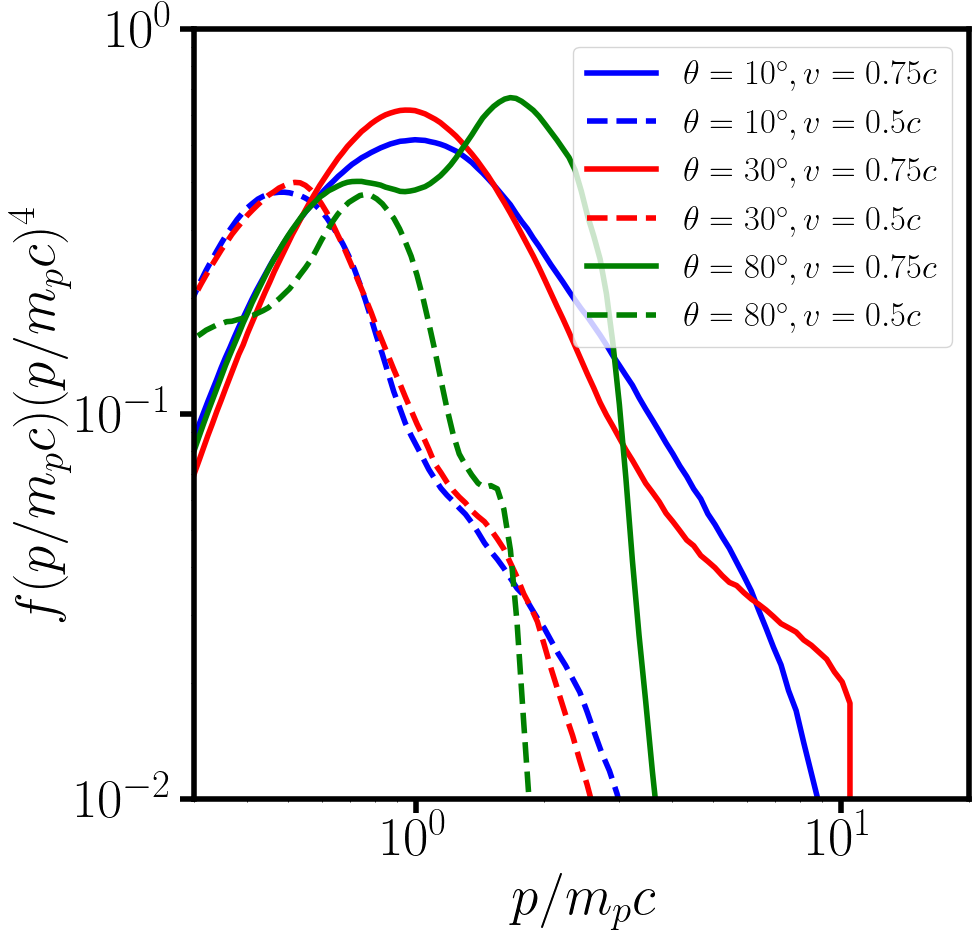} 
		\caption{Proton distribution functions in the downstream of the shocks with different inclination angles $\theta$ and velocities $v$}
		\label{distributionsP}
\end{figure}

To obtain the distributions of accelerated particles we used two approaches --- PIC simulations with code SMILEI developed by \cite{Derouillat} for low energies, and Monte Carlo simulations for high energies. In PIC simulations the shock was initialized due to the collision of the homogeneous plasma flow  with the reflecting wall. We used two values of flow velocity $v = 0.5~c$ and $v = 0.75~c$ in different setups. The magnetization was $\sigma = B^2/4 \pi m_p n \gamma c^2 = 0.004$, where $m_p$ is proton mass and $\gamma$ is Lorentz-factor of plasma flow and $n$ is the electron concentration of plasma. Simulations used 2d3v scheme, in which simulation region is two-dimensional, while particle velocities and the electromagnetic field are represented by full 3D vectors. Number of grid cells in direction of flow velocity is $Nx = 256000$ and in transverse direction $Ny = 400$. The spatial grid step is $dx = 0.2c/\omega_e$ and the time step is $dt = 0.1{\omega_e}^-1$, where
$\omega_e$ is electron plasma frequency $\omega_e = \sqrt{4\pi n e^2 / m_e}$, and $e$ is the absolute value of the electron
charge. Electron number density $n$ in units of the code is equal to the unity, and all obtained quantities can be easily scaled for different values of number density in CGS units. Electron mass is increased up to $m_e = m_p/100$ to decrease gap between proton and electron scales for economy of computational resources.  Different inclination angles of the magnetic field to the flow velocity were studied. The obtained particle distributions in the shock downstream for different velocities and magnetic field inclination angles are presented in Figure \ref{distributionsE} (for electrons) and Figure \ref{distributionsP} (for protons). One can see the well-known effect that the  efficiency of particle acceleration strongly depends on the shock obliquity \citep{SironiSpitkovsky2009pair, Caprioli2014, Crumley2019, GuoSironi2014_1, Romansky2018}.

PIC simulations are very limited in spatial and time scales, so they cannot provide particle distributions wider than a few decades. We used Monte Carlo simulations with physical parameters similar to used in PIC to obtain particle distributions on high energies. The Monte Carlo model was successfully applied to study particle acceleration by non-relativistic flows in the Earth bow shock \citep{Ellison90} and in supernova remnants \citep[see e.g.][]{EB11}.  
The stationary plane-parallel self-consistent Monte Carlo model describing particle acceleration by mildly relativistic shock, developed in \cite{BykovJETP2022}, was used to simulate particle acceleration to high energies. The model is developed for the quasi-parallel shock. In this model, during the iterative process, the conservation laws of energy and momentum near the shock are fulfilled. The model takes into account the magnetic field amplification due to plasma instabilities driven by the anisotropy of the accelerated particle distribution in the shock upstream. The Bell's  and resonant instabilities are included in these calculations. In the calculation performed on the basis of the model, the obtained ratio between the magnetic field pressure downstream and the flux parameters far upstream is

\begin{equation}\label{epsilon_B}
\epsilon_{Bd}=\frac{B^{2}_{d}}{8\pi\gamma^{2}_{0}u^{2}_{0}m_{p}n_{0}},
\end{equation}
where $B_{d}$ is the magnetic field downstream, $\gamma_{0}$ is the shock Lorentz-factor, $u_{0}$ is the shock speed, $n_{0}$ is the number density of background protons far upstream (in the background plasma rest frame system).
The setups are described in details in \cite{BykovJETP2022}. Simulation showed that in mildly relativistic shock with $\gamma_0 \approx 1.5$ protons can be accelerated to energies $\sim$ $10^8~\rm GeV$ and $\epsilon_{Bd}\approx 0.04$ which is close to the value obtained in the shock downstream in PIC simulations, and it shows that distributions obtained in two different simulations can be used together consistently. The value of $\epsilon_{Bd}$ about a few percent is typical for Monte Carlo simulations of shocks, as it is shown in our previous works \cite{BykovJETP2022, Bykov2014, BykovUniverse}. It's value slightly increases with shock velocity and reaches the plateau on relativistic velocities. The value of maximum energy is high due to the long distance to particle free escape boundary (FEB), used in the Monte Carlo simulation, which is $10^{17}~\rm cm$ or $0.5$ of the estimated radius of the source. $L_{FEB}$ is the distance from the shock to the FEB. The value of maximum energy is proportional to composition $L_{FEB}B_{st0}$, where $B_{st0}$ is the value of the RMS turbulent field in the far upstream unperturbed by CRs. In the model the turbulent magnetic field $B_{d}$ in the downstream is much higher then $B_{st0}$ because it was further amplified by plasma instabilities both in the shock precursor and downstream. This value should be considered as an upper limit for the energy of accelerated protons similar to one given by the Hillas limit. If we moved the FEB position at $0.1$ of the source's radius, then the maximum energy of a proton would decrease proportionally to the value $\sim 10^7~\rm GeV$, still above the knee in the galactic CRs spectrum. 

\begin{figure}
		\centering
	    \includegraphics[width=0.4\textwidth]{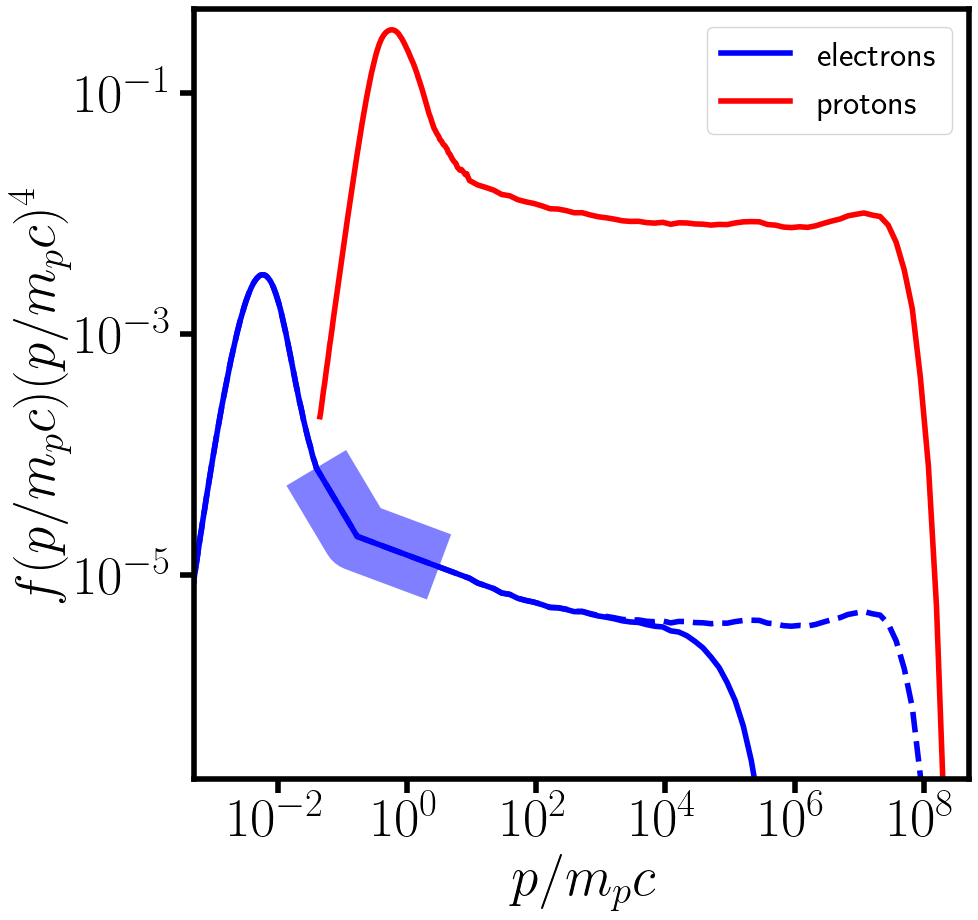} 
	    \caption{Distributions of protons and electrons, obtained with combining PIC simulation with $\theta=30^\circ$ and quasi-parallel Monte Carlo simulation with parameters $v=0.75~c$. Dashed line represents the electron distribution without taking into account synchrotron losses. Wide blue line represents the region of matching of two electron distributions (PIC and MC).}
	    \label{distributions}
\end{figure}

Monte Carlo simulations treat particle injection into the acceleration process by assuming a recipe for particle scattering rates. Here we used the microscopic description of the electrons and ions kinetics in the moderately relativistic shock obtained in PIC simulations, which provided the low energy parts of particle distribution functions (limited by the available dynamical range of PIC). Then the simulated low energy particle distributions with correct normalization were injected into Monte Carlo model of the diffusive shock acceleration to derive the high energy parts of electron and proton distributions. Diffusive approximation supposes that the distribution function for high energies is determined by the diffusion coefficient, which depends only on particle's gyroradius, and consequently it is equal for electrons and protons with same energies. In this regime  the distributions of protons and electrons are similar unless radiative losses of electrons become significant. One can find the maximum energy of electrons in the model comparing the synchrotron-Compton radiation losses $dE_{synch}/dt$ to the energy gain rate $dE_{acc}/dt$. The acceleration time of a particle from initial momentum $p_0$ to momentum $p$ in the diffusive shock acceleration can be estimated as 

\begin{equation} \label{accTime}
    t_{acc}=\int_{p_0}^p \frac{3}{v_1-v_2}\left(\frac{D_1}{v_1}+\frac{D_2}{v_2}\right)\frac{dp}{p}
\end{equation}
where $D_1, D_2$ are diffusion coefficients, $v_1, v_2$ --- flow velocities in the shock upstream and downstream, respectively \citep{Drury83}. For a relativistic particle, accelerated by a strong shock in the Bohm regime with $D = cE/3eB$, one can get the expression for energy gain rate $dE_{acc}/dt = 3 v^2 e B/8c$. Then equating this to the synchrotron losses rate, one can obtain the maximum possible energy of electrons $E_{max}= (v/c) \cdot m^2 c^4/\sqrt{e^3 B}$. In our setups the magnetic field was $B\approx0.3~\rm{G}$, as it will be described below, and the corresponding value of $E_{max}$ is $\approx10^8~m_e c^2$. So we multiplied the electron distribution function by the factor $\exp(-E/E_{max})$ to take into account the radiative losses. The distribution functions of protons and electrons in the downstream close to the shock with flow velocity $v = 0.75~c$ and magnetic field inclination angle $\theta = 30^\circ$ are presented in Figure \ref{distributions}. One can see that the  relativistic shock with $\gamma_{0} \sim 1.5$ can accelerate protons up to $10^8~\rm GeV$ and electrons to $10^4~\rm GeV$ (because of the strong synchrotron losses). Also, the injection rate of electrons is much lower than that of protons by factor $\sim 2\times10^3$ and it corresponds with the results obtained by \cite{ParkElectronIonShock}. There are some uncertainties in the matching the electron  distribution function simulated with PIC with that simulated with Monte Carlo technique. This may affect somewhat the resulting fluxes of the synchrotron radiation.

Let us estimate if such the sources can produce a  population of CRs with energies above 10 PeV in the Galaxy. The energy contained in $10 - 100$ PeV particles in the presented source model (with the hard spectrum of accelerated high energy CRs) $99$ days after the explosion is $\sim 5\times10^{48}~\rm erg$. The source likely stays mildly relativistic for a longer time, about a year, until it slows down, so the total amount of energy converted to particles with energy $>10$ PeV is several times higher. Therefore, the estimate of the energy density of the energetic particles from one such source is $\sim 10^{-19}~\rm erg~cm^{-3}$, assuming PeV regime CRs are uniformly filling the Milky Way halo  \citep{StromgMoskalenkoPtuskin2007}.
The observed energy density of CRs in the range $10 - 100~\rm PeV$ is $\sim 10^{-17}~\rm erg~cm^{-3}$. With an assumed diffusion coefficient $\sim 10^{30}~\rm cm^2~s^{-1}$ for 10 PeV particle \citep{GiacintiDiffusion, BykovKalyashovaAnisotropy}, its lifetime in the Galaxy during the diffusive propagation is $3\times 10^5$ years. This implies that the sources with mildly-relativistic shocks can provide a significant part of the observed population of 10 PeV particles in the Galaxy if the frequency of events one per 3,000 years. High energy CR protons accelerated in FBOTs and the other transient sources like the tidal disruption events \citep[see e.g.][]{KM20} can contribute into the observed extragalactic neutrino flux \citep{KM22}.

\section{Radio emission}\label{synchrotronChapter}
The observed radio spectral flux from the fast transient source CSS161010  is consistent with the synchrotron self-absorption model spectrum \citep{Coppejans2020}. The analytical estimates of the FBOT source parameters can be obtained using the synchrotron self-absorption model developed by \citet{Chevalier1998, ChevalierFransson}. This model was used to analyze the radio spectra of a number of FBOT type sources by \cite{Ho2019cow, Ho2020koala, Coppejans2020, Ho2021at2020}. Assuming a simplified cylinder geometry of the source with the axis along the line of sight and power-law distribution of electrons, and introducing the standard dimensionless parameters $\epsilon_e, \epsilon_B$ for the fractions of energy in accelerated particles and magnetic field, respectively (following the model by \cite{Chevalier1998}), one can calculate the magnetic field and the radius of the source as:  

\begin{equation}
R = {\left( \frac {6 \epsilon_B {c_6}^{s+5}{F_{max}}^{s+6}{
D}^{2s+12}}{\epsilon_e f \left( s-2 \right) {\pi}^{s+5}{{c_5}}^{s+6}
{E_1}^{s-2}} \right)} ^{ \frac{1}{2s+13} } \frac{2 c_1}{\nu},
\end{equation}

\begin{equation}\label{ChevB}
B = { \left(\frac {{\epsilon_B}^2 36 {\pi}^{3}{c_5}}{{
\epsilon_e}^{2}{f}^{2} \left( s-2 \right) ^{2}{{c_6}}^{3}{{E_1}}^{2 s-4}F_{max}{D}^{2}} \right)}
^{\frac{2}{2s+13}}\frac{\nu}{2 c_1},    
\end{equation}
where $F_{max}$ is maximum spectral flux density at the given time moment, $\nu$ is its frequency, $\displaystyle s$ is the electron spectral index , $\displaystyle$ $c_1$, $c_5$, $c_6$ are the Pacholczyk constants~\citep{Pacholczyk}, depending on $\displaystyle s$, $\displaystyle E_{1}$ is the minimum energy of
the electron power law distribution, $\displaystyle f$
is the filling factor---the emitting fraction of the source volume, and~$\displaystyle D$ is the distance to the source. After calculating the source radius  one also can obtain the average transverse velocity of expansion, by dividing radius by time since explosion.
Parameters, obtained by \cite{Coppejans2020} in equipartition regime for time moment $t_{obs} = 99$ days after explosion, are: $\epsilon_e = \epsilon_B = 1/3$, spectral index $s = 3.5$, source radius $R = 1.4\times10^{17}~\rm cm$, magnetic field $B = 0.29~\rm G$, number density $n = 25~\rm{cm^{-3}}$, velocity $\Gamma\beta = 0.55$ and corresponding total energy $5.6\times10^{49}~\rm erg$. 

In this work we suggest the improvement of this approach. To calculate synchrotron radiation, we used electron distribution function obtained from PIC simulations of shock wave instead of power-law distribution, and this allowed us not to use arbitrary parameters $\epsilon_e$, $s$ and $E_1$. With this distribution function we calculated synchrotron radiation using the numerical integration of standard formulae for emissivity and absorption coefficient, described in \cite{Ginzburg1975, Ghisellini}. We chose values of the magnetic field and the number density to fit the observational data. Emissivity per unit frequency per unit volume is

\begin{equation} \label{emission}
I(\nu)=\int_{E_{min}}^{E_{max}} dE \frac {\sqrt {3}{e}^{3}n f(E) B_{\perp}}{{m_e}{c}^{2}}
\frac{\nu}{\nu_c}\int_{\frac {\nu}{\nu_c}}^{\infty }\it K_{5/3}(x)dx,
\end{equation}
where $f(E)$ is electron distribution function, $B_{\perp}$ is the magnetic field component normal to the line of sight, $m_e$ - mass of electron and $\displaystyle\nu_{c}$ is the critical frequency
$\displaystyle\nu_{c} = 3 e^{2} B_{\perp} E^{2}/4\pi {m_{e}}^{3} c^{5}$, and~$K_{5/3}$ is
the modified Bessel function. The~absorption coefficient is

\begin{equation}\label{absorption}
k(\nu)=\int_{E_{min}}^{E_{max}}dE\frac{\sqrt {3}{e}^{3}n B_{\perp}}{8\pi m_e \nu^2E^2}
\frac{d}{dE} E^2 f(E)\frac {\nu}{ \nu_c}\int_{\frac {\nu}{ \nu_c}}^{\infty }K_{5/3}(x) dx.
\end{equation}

We assumed the source is a cylinder with radius $R$, and height $h = f\cdot R$, with parameter filling factor f, and axis directed to the observer. Also for simplicity we assumed that number density, magnetic field and electron distribution are homogeneous inside the source. After the numerical integration of formulae \ref{emission} and \ref{absorption} through the line of sight, we obtained the radio flux for given magnetic field and number density.

\renewcommand{\arraystretch}{1.375}
\begin{table}[h!]
		\label{parameters}
		\caption{Obtained parameters of radio source for different models}
		\begin{center}
			\begin{tabular}{|c | c| c| c| c| c|}
                 \hline
                 model &$r$ & $B$, G& $n$,~$\rm cm^{-3}$&  $E_e,~\rm erg$& $E_p,~\rm erg$\\
                 \hline
                 A80 & 550 & 0.5 & 2$\times 10^6$  & 3.4$\times 10^{51}$ & 3$\times 10^{53}$\\
                 \hline
                 A30 & 140 & 0.1& 5$\times 10^5$&  8$\times 10^{51}$ & 7$\times 10^{53}$\\
                 \hline
                 B80 & 170 & 0.3& 540&  3$\times 10^{49}$& 3.7$\times 10^{51}$\\
                 \hline
                 B30 & 57 & 0.34& 17&  2.2$\times 10^{48}$& 1.4$\times 10^{50}$\\
                 \hline
		\end{tabular}
        \end{center}
\end{table}

Fitting the data and evaluating magnetic field and number density was performed with minimization of the functional
$\displaystyle r(B,n) = \sum (F(\nu_{i}, B, n) - F_{obs}(\nu_{i}))^{2}/\sigma_{i}^2$, where $\nu_i$ 
are the observed frequencies, $\displaystyle F_{obs}(\nu_{i})$ are the observed fluxes, $\sigma_{i}$ - observational errors and
$\displaystyle F(\nu_{i}, B, n)$---the modeled fluxes. Radius of the source was considered fixed for each chosen velocity of the source expansion $R = v\cdot t_{obs}$ and filling factor is $f = 0.5$. The data fitting results for models with different electron distribution functions are presented in Table 1. Here $r$ is the resulting residual, $B$ - magnetic field, $n$ - number density of electrons, $E_e$ is the kinetic energy of electrons and $E_p$ is the kinetic energy of protons, calculated with integration of distribution functions. Models A30 and A80 use distributions from PIC simulations with shock 
 velocity $v = 0.5~c$ and magnetic field inclination angles $30^\circ$ and $80^\circ$ respectively, and Models B30 and B80 use distributions from simulations with $v = 0.75~c$. One can see, that the models with velocity $0.5~c$ have much worse residuals and also too high number density, which leads to unphysical amount of released energy. Models with higher shock velocity provide more reasonable results. Obtained magnitude of the magnetic field and the number density are close to the values provided in \cite{Coppejans2020}, while the derived velocity estimation is about $0.75~c$  which is significantly higher than their estimated value of $0.5~c$. The difference is because we are using the distribution function of relativistic electrons and $\epsilon_e$ parameter from our PIC model which are different from that were  postulated by \cite{Coppejans2020}. The distribution function which we obtained with the PIC model to explain the observed radiation for $v = 0.5~c$ requires apparently unreasonable high number density. Also, the obtained high shock velocity value is supported by the result of the analysis of AT2018cow by \cite{BietenholzNoJet}, where the direct velocity estimation revealed a substantially higher value, than that was obtained with the standard synchrotron self-absorption technique assuming the postulated default choice of $\epsilon_e$ and the other parameters \citep{Ho2019cow}.
 
\begin{figure}
		\centering
	    \includegraphics[width=0.4\textwidth]{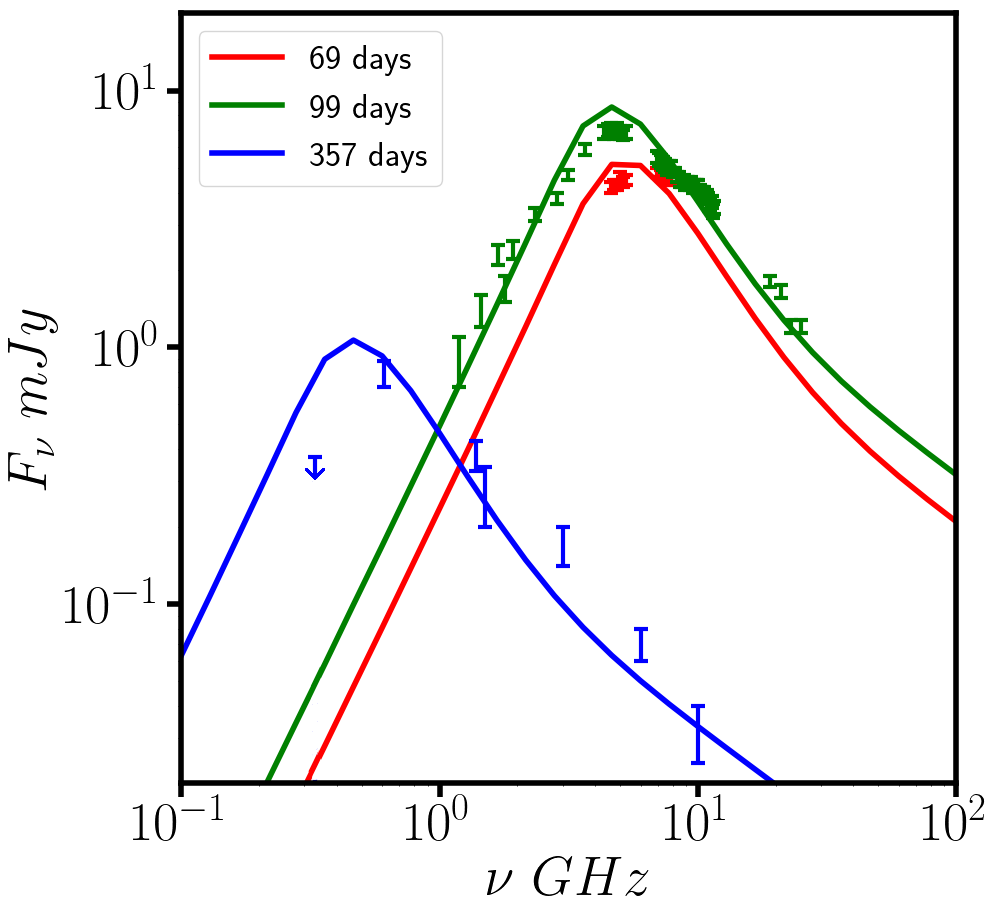} 
	    \caption{Observed and modeled (for model B30) radio spectra of CSS161010 at 69, 99 and 162 days after the explosion.}
	    \label{synchrotron}
\end{figure}

Modeled radio flux for the shock with flow velocity $v = 0.75 c$ and magnetic
field inclination angle $\theta  = 30^\circ$ (model B30) and observational data for different time moments are shown in Figure \ref{synchrotron}. The model fits for different time moments were performed independently, obtained parameters for the moment 99 days after explosion are listed in Table 1 while for the rest two fits these are the following. At 69 days the magnetic field equals $0.41~\rm G$, the concentration $24~\rm cm^{-3}$ and radius of the source is $R =1.3\times10^{17}~\rm{cm}$. At 357 days the magnetic field is equal $0.04~\rm G$, the concentration is $0.25~\rm cm^{-3}$ and the source radius is $R= 6.9\times10^{17}~\rm{cm}$. The obtained values of magnetic field and number density are significantly differ from values expected under assumption that $B \propto r^{-1}$ and $n  \propto r^{-2}$. Observation at 162-nd day after explosion can not be fitted adequately with our model. It requires the shock wave velocity and electron distribution function different from our shock propagation model. It could be explained by the strong macroscopic inhomogeneity of the circumstellar medium in the vicinity of the shock. The peculiarity of this observation was first noticed by \cite{Coppejans2020}.

Our modeling, as well as the analysis provided by \cite{Coppejans2020} using Chevalier's method, shows that the magnetic field in the system should be rather strong (about $0.3~\rm G$ at distances about $10^{17}~\rm cm$). It does not seem to be possible that quasi-uniform interstellar magnetic field can be so strong. Monte-Carlo modeling provides amplification of magnetic field to close values, but the modeling of radio emission was independent from Monte-Carlo. With obtained parameters, radiation is produced by the electrons with Lorentz-factors about several hundreds. Long electron distributions, and consequently strong amplification of magnetic field, are not necessary for modeling of radio emission. So another option is to use the magnetic field generated by the star, but on these distances it should be perpendicular to the shock velocity \citep{Parker} and particles cannot be accelerated in quasi-perpendicular shocks as it was mentioned above. The presence of strong turbulence can increase the efficiency of acceleration in quasi-perpendicular shocks, as it was shown with PIC simulations in \cite{Demidem2023inhomogenousshock, Bresci2023turbulentchock, Romansky2019turbulence}. The strong magnetic field and high density may exist inside massive star clusters, as shown by MHD modeling in \cite{BadmaevCluster}.
 
\section{X-ray radiation}\label{compton}

While radio emission from CSS161010 can be explained with the model of quasi-spherical shock with synchrotron radiation and self-absorption, the bright X-ray radiation detected on the 99 day after the explosion (with flux $F_{X} = (1.33 \pm 0.76)\times10^{-15} ~\rm erg~s^{-1}~cm^{-2}$ and corresponding isotropic luminosity $L_{X}=(3.4\pm1.9)\times10^{39}~\rm erg~s^{-1}$ in the range $0.3-10~\rm keV$ \cite{Coppejans2020}) is not easy to explain. Recent studies of FBOTs use central engine to explain powerful X-ray emission from them --- on the early stages it is produced by millisecond magnetar \citep{GottliebFBOTjet} and on the late stages by the accretion disk  \citep{MiglioriFBOtlatestages}. Here we present possible models of X-ray emission form FBOT (particularly CSS161010) without using central engine.

Electrons with distribution obtained from PIC simulations cannot produce such strong X-ray radiation even with prolonged to the high energies power-law tail with spectral index $\sim 3.5$. But as Monte-Carlo simulation showed, the electron distribution becomes harder and on high energies it has spectral index $\sim 2$ (see Figure \ref{distributions}). Using the  composite distribution function obtained in section \ref{particlesChapter} and the magnetic field $B \approx 0.34~\rm G$ from model B30 in section \ref{synchrotronChapter}, one can calculate the synchrotron X-ray flux from this source.  The lifetime of emitting electrons should also be taken into account. The energy of electrons emitting photons with energy $E_{ph}$ in the magnetic field $B$ is $E_e = (4\pi{m_e}^3 c^5 E_{ph}/0.29\cdot3 e B h)^{1/2} \approx 2\cdot10^6~m_e c^2$ if we take $E_{ph} = 5~\rm keV$. The lifetime of electrons with this energy is $\tau = {m_e}^4 c^7 / e^4 B^2 E_e \approx 4\times10^3~\rm s$. The distance that electron can reach by this time, moving diffusively in the shock downstream  is  $d \approx 2\times10^{13}~\rm cm$. This means that the emitting volume drastically decreases by factor of $10^4$ and we should calculate the radiation of layer with width $d$. Thus the estimation of the resulting  synchrotron X-ray flux is $10^{-16}~\rm erg~s^{-1}~cm^{-2}$, which is much smaller than the observed. But the model B30 has arbitrary parameter --- filling factor $f=0.5$, while the synchrotron spectrum, obtained as described in section \ref{synchrotronChapter}, depends on the product of filling factor and electron number density. If we use thinner but more dense shell, with filling factor $0.1$ and electron number density $80~\rm cm^{-3}$, the effect of decrease of the emitting volume due to synchrotron energy losses at high energy becomes weaker. Estimates based on the expression \ref{epsilon_B} and the model parameters B30 with 
the electron number density $80~\rm cm^{-3}$  give the value of the magnetic field in the downstream $B\approx 0.2$ G in Monte Carlo calculations which within the errors of the models agrees well with the found value. The standard hydrodynamical Sedov's solution supposes that the width of the expanding shell is $1/12$ of its radius, and the shocks with significant fraction of accelerated particles can have stronger compression ratio \citep{Toptygin1997} and consequently be thinner. That is why we consider the filling factor of $0.1$ which is possible for the mildly-relativistic shock model.  

\begin{figure}
		\centering
	    \includegraphics[width=0.4\textwidth]{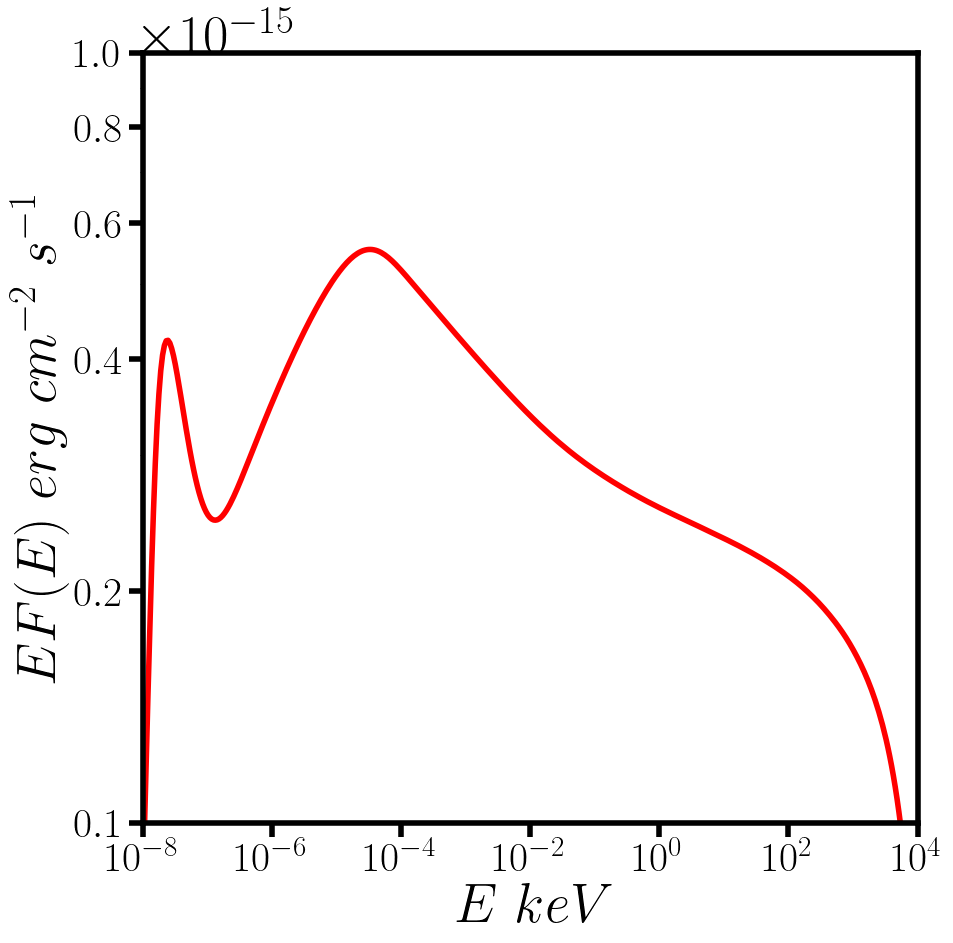} 
	    \caption{Modeled synchrotron spectrum from the electron distribution obtained in section \ref{particlesChapter}. Integrated X-ray flux in the range $0.3-10~\rm keV$ is $8.9\times10^{-16}~\rm erg~cm^{-2} s^{-1}$ in the model. It is consistent within the errors with the X-ray flux detected by Chandra $F_{X} = (1.33 \pm 0.76)\times10^{-15} ~\rm erg~s^{-1}~cm^{-2}$. The energy flux curve was derived with the model where the ambient electron density was deduced from the radio flux model given in Fig.\ref{synchrotron}.}
	    \label{long_radiation}
\end{figure}

For the numerical modeling of X-ray synchrotron spectrum we used the source with the width $\sim 0.1$ of its radius (which is $R=1.9\times10^{17}~\rm cm$) and the electron number density $80~\rm cm^{-3}$. The electron distribution function in every point depends on the distance to the shock front due to the synchrotron losses. At the shock front we used the distribution $f(E)$, shown in Figure \ref{distributions} while at the distance $l$ behind the shock we used:

\begin{equation}
    f_l(E)=f\left(\frac{E}{1-4e^4 B^2 E~l/9m^4 c^7 u_d}\right)\cdot\frac{1}{\left(1-4e^4 B^2 E~l/9m^4 c^7 u_d\right)^2}
\end{equation}
where $u_d$ is the shock downstream velocity, which equals $0.2~c$ in our setups with initial flow velocity $v=0.75~c$. The resulting spectrum of the source in the wide energy range is shown in Figure \ref{long_radiation}. The first peak of the curve is produced by the synchrotron self-absorption, the dip and the following increasing part of the spectrum are due to the hardening of the electron distribution (shown in Figure \ref{distributions}). The second spectral peak and the following fading are due to the synchrotron losses of the relativistic electrons. The photon index at the high energies is approximately $2$ which is close to the value obtained in \cite{Ho2023flares} for the transient source AT2022tsd. The total flux in the range $0.3-10~\rm keV$ is $8.9\times10^{-16}~\rm erg~cm^{-2} s^{-1}$. This value is consistent with the observed X-ray flux from CSS161010 at 99 days after the explosion.

To summarize the properties of our model let us estimate the parameters for the energy and mass balance. The total kinetic energy in the emitting shell is determined by protons and it is equal to $1.4\times10^{50}~\rm erg$ as it is shown in the Table 1. The energy in electrons is $2\times10^{48}~\rm erg$ and the energy in magnetic fields is about $10^{49}~\rm erg$. The magnetic field in our radiation model corresponds well to that obtained in Monte Carlo simulations with account for the magnetic field amplification. The total mass of the emitting shell is $3\times10^{29}~\rm{g} \approx 10^{-4}~ M_\odot$, and it originates from the surrounding matter swept-up by the forward shock. The mass of the ejected matter should be several times higher in order to  the shock not to slow down till the given time moment. The mass of $\sim 10^{-3}~ M_\odot$ moving with total energy less than $10^{51}~\rm erg$ is a reasonable estimate for the relativistic ejecta. We conclude that the parameters of our model describing X-ray radiation from FBOT CSS161010 via the synchrotron radiation seems to be realistic and consistent with the observational data.

\section{Jet - massive star interaction model}\label{jetChapter}
FBOT model presented by \citet{GottliebFBOTjet} suggested a presence of a 
powerful jet produced by a central engine. Interaction of the jet with the surrounding matter and stars can be a source of non-thermal radiation as well. While the mildly relativistic quasi-spherical shock model presented above can explain the non-thermal X-ray  flux, let us consider a possibility of an explanation of the observed X-rays with the jet model. The interaction of a jet with a nearby star can produce  the X-ray emission via inverse Compton (IC) scattering of accelerated electrons on the photons of this star. This assumption is supported by studying of the local environment of object AT2018cow by \cite{SunAT2018environment}, showing that there are two young massive star clusters in the vicinity of AT2018cow. Young massive star clusters (YMSCs) have a lot of stars with strong winds and high luminosity in a rather small volume (e.g. cluster Westerlund 1 has 6 yellow giants, 4 red supergiants, 24 Wolf-Rayet stars, and dozens of OB stars in the radius of 1 parsec \citep{Clark2005westerlund, Crowther2006westerlund, Negueruela2010westerlund}. So if the supernova explosion happens in YMSC, it is possible that the ejecta will interact with the strong wind of neighbour star, placed at distance $\sim 10^{17}~\rm cm$.

The common approach to estimate the X-ray luminosity, knowing the radio luminosity, uses well-known formula for the ratio of luminosities of inverse Compton and synchrotron radiation from the same electrons, which equals to the ratio of the photon and  magnetic energy density $L_{IC}/L_{synch}=u_{ph}/u_{B}$ \citep{Ghisellini}. However, this expression does not take into account synchrotron self-absorption, which is very important for FBOTs radiation. For the model B30 the total radio luminosity computed without self-absorption is 5 times greater than the actual luminosity, and for the model B80 it is 20 times greater. So synchrotron self-absorption makes a restriction on energy densities ratio weaker on a factor about 10. The ratio of radio and X-ray luminosities at 99-th day after the explosion is $L_{X}/L_{radio} \approx 10$ and the magnetic energy density in CSS161010 is $u_{B}=B^2/8\pi \approx 0.003~\rm erg~cm^{-3}$. So to explain the X-ray flux as IC radiation of the same electrons that produce synchrotron radio flux, we need photon energy density $\sim 0.003~\rm erg~cm^{-3}$. It is much greater than energy density of CMB $\sim 4\times10^{-13}~\rm erg~cm^{-3}$ and mean galactic photon field $\sim 10^{-12}~\rm erg~cm^{-3}$ \citep{Mathis1983} and also greater than the most luminous stars with $L \approx 10^{40}~\rm erg~s^{-1}$ can provide on distances about supposed size of radio-emitting shell $\sim10^{17}~\rm cm$: $L/4\pi R^2 c \approx 10^{-6}~\rm erg~s^{-1}$. So we consider that some close energetic source of initial photons is necessary for explanation of X-ray radiation.

\begin{figure}
	\centering
	\includegraphics[width=0.49\textwidth]{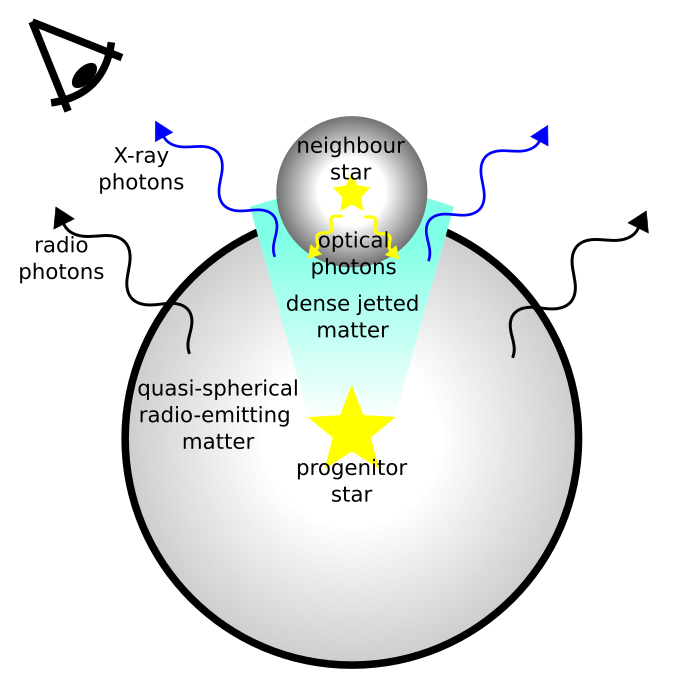} 
	\caption{Scheme of emission model. Quasi-spherical and relatively sparse fraction of shocked matter emits in radio while dense jet emits X-rays via IC scattering of optical photons from close neighbour star.}
	\label{scheme1}
\end{figure}

 One can consider a model which is based on IC scattering of photons from a close neighbour star by the electrons accelerated in the expanding FBOT remnant. Let us estimate the necessary parameters of such a system. The X-ray measurements were obtained for the energy range $0.3-10~\rm keV$, so we consider the X-ray photons of energy $E_x  \sim 1~\rm keV$. The seed photons are provided by a hot luminous star with the energy $E_{ph} \sim 1~\rm eV$. To convert the seed optical photons to X-rays, we need relativistic electrons with the moderate Lorentz-factors  $\gamma \sim \sqrt{E_x/4E_{ph}} \sim 15$. The energy density of the seed optical photons is $U_{ph} = L_0/4\pi R_x^2 c$, where $L_0$ - luminosity of the neighbour star and $R_x$ - distance to the star from the X-ray source, which is assumed to be a cylinder with radius $R_x$ and height $0.5 R_x$. The total IC luminosity is equal to
 
\begin{equation}
    \label{comptonluminosity}
    L_{IC}=\frac{4}{3}\sigma_T c \gamma^2 U_{ph} n_e 0.5 \pi R_x^3
\end{equation}
where $\sigma_T$ is the Thompson cross-section and $n_e$ is the number density of emitting electrons. Assuming that the X-ray radiation originates from IC scattering of the seed photons from a high luminosity star with  $L_0 = 2\times10^{39}~\rm erg~s^{-1}$ (e.g. a Wolf-Rayet star) one can calculate the necessary energy of electrons, their number density and the radius of the source. 
The obtained value of radius of X-ray source is  $R_x\approx2\times 10^{15}~\rm cm$. This is much smaller than the radius of radio-emitting shell calculated above, so we need another population of electrons to explain the X-ray radiation with IC scattering. 

The required  population of the energetic electrons can be provided by a jet-type outflow from the progenitor star explosion, moving in the direction towards a close neighbour star. This division of FBOT's ejecta in two populations corresponds with studying of GRB jets, which show structured anisotropic distribution of the  ejected energy \citep{GranotJets, GottliebJets, SalafiaJets2}. The sketch illustrating this model is presented in Figure \ref{scheme1}. At the moment of time $t_{obs}$ the jet can extend to the distance $d\sim c\cdot t_{obs}\approx2.5\times10^{17}~\rm cm$. With the found value of $R_x$ we can estimate that the jet opening angle $\theta = R_x/d \approx 0.01$. Typical long GRB jets have the opening angles $\sim 10^{\circ}$ or $0.2~\rm rad$ the angles $\sim 0.01~\rm rad$ are rare but still possible \citep{GoldsteinJetAngles, LloydRonningJetAngles}. The number density of electrons estimated in this scenario  $n_e\approx10^7~\rm cm^{-3}$ seems to be very large. Strong winds from Wolf-Rayet stars with mass-loss rate $\sim 10^{-4}M_\odot$ per year and wind velocity $\sim 1000~\rm km/s$ can provide density $\sim 10^6~\rm cm^{-3}$ at distances of $\sim 10^{15}~\rm{cm}$. The ratio of the proper densities at the strong relativistic shock is $(\hat{\gamma} \gamma -1)/(\hat{\gamma} -1)$ where $\gamma$ is Lorentz-factor and $\hat{\gamma}$ --- adiabatic index of matter behind the shock as it is shown by \cite{Blandford1976}. In the lab frame this ratio increases by factor $\gamma$ due to the relativistic compression. Propagation of GRB jets in dense wind medium was studied in \cite{GranotJets,MeslerJetDense}, and it shows that the jet can continue propagation and produce afterglow for months after the initial explosion. Some numerical MHD simulations \citep{Granot2000, Granot2001} show even higher density in the thin jet head in 100 days after explosion.  Also theoretical studying of GRB jets shows \citep{KumarGranot} that the ejected matter is compressed to the thin shell with thickness $R/4\gamma^2$, where $R$ is the distance from the jet head to the progenitor star. Such strong compression is suitable for our assumption of a dense emitting shell.

\begin{figure}
	\centering
	\includegraphics[width=0.49\textwidth]{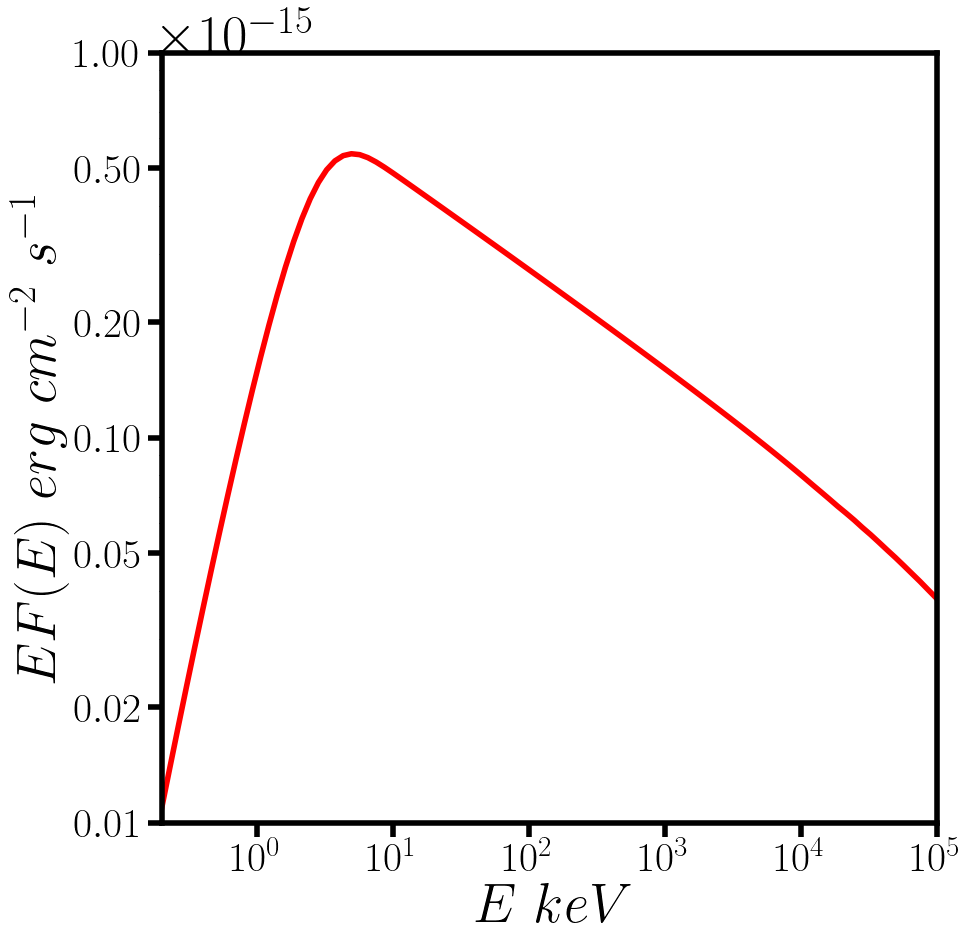} 
	\caption{Modeled spectrum of inverse Compton radiation produced by interaction of electrons accelerated in jet shock with photons from close neighbour star.}
	\label{compton_radiation}
\end{figure}

We performed the numerical computations of the X-ray radiation produced via IC scattering using self-developed numerical code with the following parameters: source radius and distance from the jet head to the star are $2
\times10^{15}~\rm cm$, star luminosity is $2\times10^{39}~\rm erg~s^{-1}$ , its temperature is $2.5\times10^4 K$. We assumed the anisotropic photon distribution directed against z-axis with half-width angle $45^\circ$. Distance to the jet progenitor star is $2\times10^{17}~\rm cm$, and consequently jet opening angle is $0.01~\rm rad$. The final photons with energies in range $0.3-10~\rm keV$ are directed at angle $10^{\circ}$ from z-axis, which is greater than jet opening angle, and due to that fact the gamma-ray burst from this object was not observed. The electron number density $n_e = 1.5\cdot10^7~\rm cm$ and their distribution is a power-law function with minimum energy $10 ~m_e c^2$ and spectral index $3.5$. As it follows from the analytical approach in \cite{GranotJetAnalitic1, GranotJetAnalictic2}, a jet, propagating in strong stellar wind medium, stays relativistic on times and distances interesting for our  modeling, so we suppose shocked electrons to be highly relativistic. Spectrum of IC radiation in this model is shown in Figure \ref{compton_radiation}. The resulting X-ray flux in $0.3-10~\rm keV$ energy range is $9.4\times10^{-16}~\rm erg~s^{-1}~cm^{-2}$ which is consistent with the observed value of $F_{X} = (1.33 \pm 0.76)\times10^{-15}~\rm erg~s^{-1}~cm^{-2}$.

While the considered model can explain the detected X-ray radiation from CSS161010, it is not very suitable for the other FBOT type objects with very high and continuous X-ray luminosity from the first weeks  after the explosion like it was detected in AT2018 \citep{Ho2019cow, Margutti2019} and AT2020mrf \citep{YaoAt2020mrf}. To address the issue the model can be modified, assuming the FBOT source in a  binary star system, where the companion star highlights the jet since early stages. The advantage of the modified model is that it provides high X-ray luminosity since the early stages of jet propagation and also doesn't depend crucially on jet direction relatively to companion star. But at the late stages the photon field from close companion would not be enough to produce strong X-ray radiation as it is described above. 

\begin{figure}
	\centering
	\includegraphics[width=0.49\textwidth]{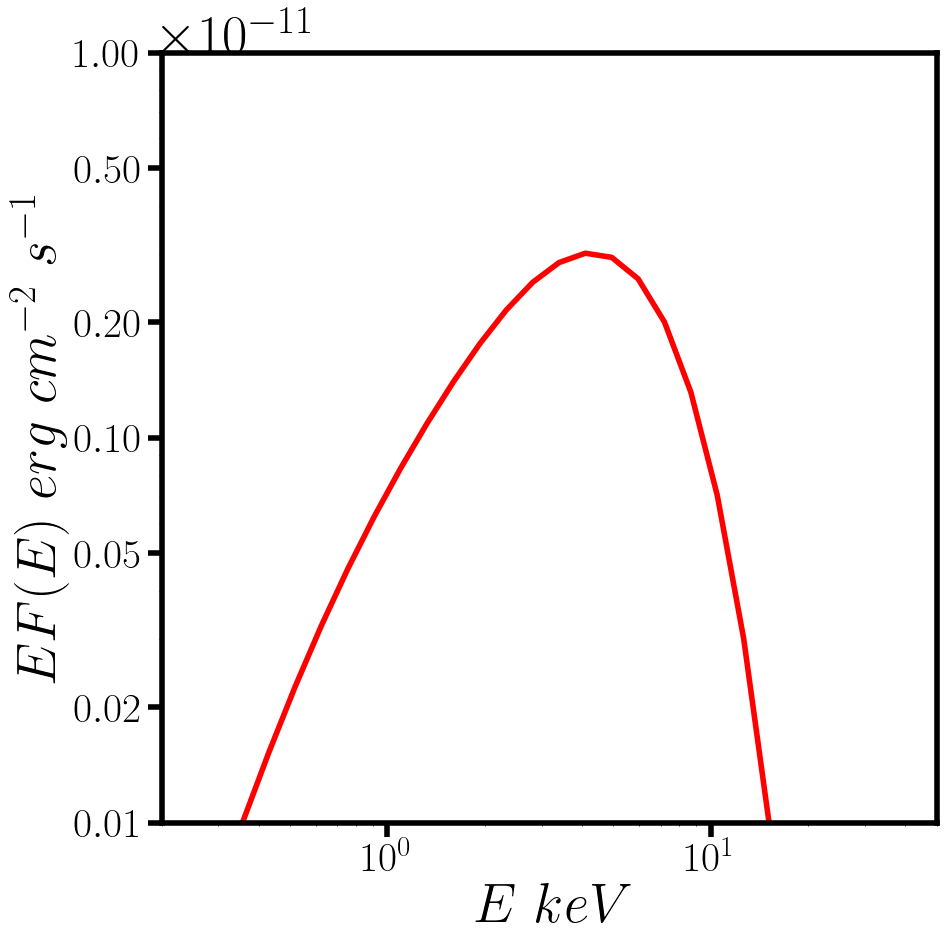} 
	\caption{Modeled spectrum of inverse Compton radiation produced by interaction of bulk electrons in jet with photons from close binary star.}
	\label{compton_radiation1}
\end{figure}

The typical distance between the companions in a binary is about $10^{14}-10^{15}~\rm cm$ \citep{RaghavanStars} so it allows to obtain very high photon density. Jets at early stages have higher Lorentz-factors, and the electron distribution in the unshocked ejecta is anisotropic, so this system can produce powerful X-ray radiation to the direction close to the jet axis via IC scattering. Particle acceleration can be efficient in relativistic outflows with both the electron-ion \citep{2023A&A...676A..23C} and electron-positron \citep{2019PhRvL.123c5101L} compositions. The details strongly depend on the model of jet propagation. For simplicity we used the following parameters: ejecta with the same mass $5\times10^{-5}M_\odot$ as in the first model, with Lorentz-factor $30$ and opening angle $5^\circ$ is placed on distance $2\times10^{14}~\rm cm$ from the companion star --- red giant with luminosity $4\times10^{37}~\rm erg~s^{-1}$ and temperature $5000~\rm K$, and the direction to the observer is $10^\circ$ from jet axis. The X-ray spectrum, obtained in numerical computations with such parameters is shown in Figure \ref{compton_radiation1}, and flux in the range $0.3-10~\rm keV$ is $3.4\times10^{-12}~\rm erg~s^{-1}~cm^{-2}$, which corresponds to isotropic X-ray luminosity $9\times10^{42}~\rm erg~s^{-1}$ in a few days after the event start. The models which combined both quasi-isotropic and jet-type outflows were discussed to describe the radiation from ultraluminous X-ray sources \citep[see e.g.][]{2022ApJ...937....5S,2024arXiv240114770C}.    

\section{Conclusions}
In this work we present a model of the origin of luminous X-ray radiation from FBOT CSS61010. We performed numerical modeling of particle acceleration in mildly relativistic shock with Lorentz-factor $\sim$ 1.5 with PIC and Monte Carlo techniques. 
Kinetic modeling provided the distribution functions of protons and electrons in the wide range of energies which we used to calculate the synchrotron radiation spectrum. The model allows to derive the standard parameters $\epsilon_{B}$ and $\epsilon_{e}$ widely used in models of the synchrotron radiation from supernovae \citep{Chevalier1998}.  Magnetic field and number density in the source were obtained by fitting modeled radio spectrum to the observational data. With these parameters we calculated X-ray flux in the range $0.3-10~\rm keV$ and obtained the value  $8.9\times10^{-16}~\rm erg~cm^{-2}$, consistent with the observed flux $F_{X} = (1.33 \pm 0.76)\times10^{-15} ~\rm erg~s^{-1}~cm^{-2}$ in  99 days after the energy release. 

We discussed two possible models of the energy release from the central source of FBOT. One with a broad shell of a mildly relativistic ejecta and another with a presence of a jet. The shell model allowed us to explain the observed non-thermal emission from object CSS161010 within a standard set of the assumptions on FBOT's circumstellar environment. On the other hand, the jet model required some fine tuning of the parameters including a very narrow opening angle of the jet to fit the data.

The high X-ray luminocity during the first weeks after the event from other FBOT-type sources, e.g. AT2018cow and AT2020mrf, which was not observed from CSS161010, can be understood in the model which accounts for the stellar source binarity. Indeed, the interaction of the jet or anisotropic relativistic outflow with the stellar companion may provide the high X-ray luminosity during the first weeks of the FBOT evolution as it was observed in AT2018cow and AT2020mrf. 

The maximum energies of CRs accelerated by the mildly relativistic shocks in FBOTs can be as high as $\sim 10^8$ GeV. With the CR spectral distribution as given in Fig. \ref{distributions} and the expected FBOT's event rate (about $\sim$ 1\% of the core-collapsed supernova rate), FBOT-type objects could contribute substantially to the observed fluxes of the high energy CRs in the Milky Way.

\section*{Acknowledgements}
The authors thank the two referees for constructive and useful comments.
Particle-in-cell plasma simulations and non-thermal radiation modeling by  V.I.R. and A.M.B. were supported by the RSF grant 21-72-20020. The modeling was performed at the Joint Supercomputer Center JSCC RAS.  The Monte Carlo modeling of mildly relativistic shock by O.S.M. was supported by the baseline project FFUG-2024-0002 at Ioffe Institute and performed at the Peter the Great Saint-Petersburg Polytechnic University Supercomputing Center.
The authors thank Dr. M. E. Kalyashova for important and useful remarks.

\bibliographystyle{model5-names}
\biboptions{authoryear}
\bibliography{fbotcompton}

\end{document}